# Planar Scale-Invariant Waveguides and Resonators with Uniform Air-Confined Modes

MOHAMMAD ENJAVI[1,†], AMIN KHAVASI[1,†], ASHKAN ZANDI[1], SAEED JAVADIZADEH[1], REZA MARZBAN[1], DEVIN K BROWN[2], ALI ADIBI[1,*]

[1]School of Electrical and Computer Engineering, Georgia Institute of Technology, Atlanta, GA, 30332 USA
[2]Institute for Matter and Systems Department, Georgia Institute of Technology, Atlanta, GA, 30332 USA
[†]These authors contributed equally to this work.
*ali.adibi@ece.gatech.edu

**We demonstrate a planar metamaterial-based resonator and waveguide with strong light confinement in air based on a silicon-on-insulator (SOI) platform that exhibits scale invariance in the lateral direction. By embedding a sub-wavelength grating (SWG) region between two silicon ridges, the waveguide maintains a nearly constant effective index across varying widths while sustaining a uniform field distribution. Simulations and experimental measurements using Mach–Zehnder interferometers confirm scale invariance, and racetrack resonators fabricated from the same structure exhibit an intrinsic quality factor of ~40,000. The ability of the resonance-based structures for confining light in air, providing large interaction regions with high quality factors along with compatibility with CMOS fabrication processes and robustness against fabrication imperfections make them excellent candidates for enhanced light–matter interaction applications with improved power handling, offering a promising platform for integrated photonics.**

Resonance-based integrated photonic structures have been extensively studied for light-matter interaction [1], and several designs have been proposed for applications like sensing [2-4], atom–photon coupling [5,6], and nonlinear optics [7–9]. Despite impressive progress in forming structures with high quality factors (Qs) or low mode volumes, confinement of resonant modes in a high-index material has been a major issue in achieving very high efficiencies due to the interaction of matter (e.g., an analyte or atomic vapor) with the evanescent tail of the optical mode. Another practical limitation is the small region of interaction for a typical whispering gallery mode resonator, which can elongate the sample delivery or the interaction time. These challenges necessitate the design of resonators with high mode confinement in the low-index material (e.g., air or water) in a large structure with a large interaction volume.

Slot resonators can partially address this need, but at the cost of very low quality factors (Qs) for large structures [10]. Although sub-wavelength-grating (SWG)-based structures can also provide reasonable Qs for smaller sizes, the nonuniformity of their mode pattern inside the low-index region reduces the efficiency of light-matter interaction [11]. To address this issue, scale-invariant devices have recently emerged; these structures can be scaled without altering their optical properties [12-16]. In cavities, this allows retention of single-mode lasing even at large sizes [13]. Extending scale invariance to waveguides has attracted attention for its ability to maintain a nearly constant effective index across varying widths [12]. Such structures support uniform, non-Gaussian field profiles in low-index regions, enhancing light–matter interactions. Uniform field distributions are also advantageous for imaging and LiDAR, where beam quality and detection accuracy are critical [17-19]. Rodrigues et al. demonstrated vertical scale-invariant waveguides using all-dielectric multilayer structures [12], but this approach is less compatible with planar integration and requires multi-material cores. Recently, waveguides with flat-top optical modes with uniform intensity profiles have been proposed [20] using silicon (Si) and silicon oxide ($SiO_2$) regions. Nevertheless, scale-invariant resonators with high field confinement in air and reasonably high Qs have not been demonstrated to date. In this paper, we demonstrate a new class of scale-invariant high-Q resonators with light confinement in air. Starting with scale-invariant waveguides formed in the Si-air platform, we will theoretically design and experimentally demonstrate scale-invariant resonators with reasonably high Qs (Q ~ 40,000), which is to the best of our knowledge, the largest value reported for such resonators with light confinement in air. These structures are compatible with widely adopted CMOS processes and achievable through lithography and a single dry-etching step. We also discuss the potential for high power handling and improved alignment tolerance, underscoring the advantages of our scale-invariant waveguide/resonator platform for integrated photonic applications.

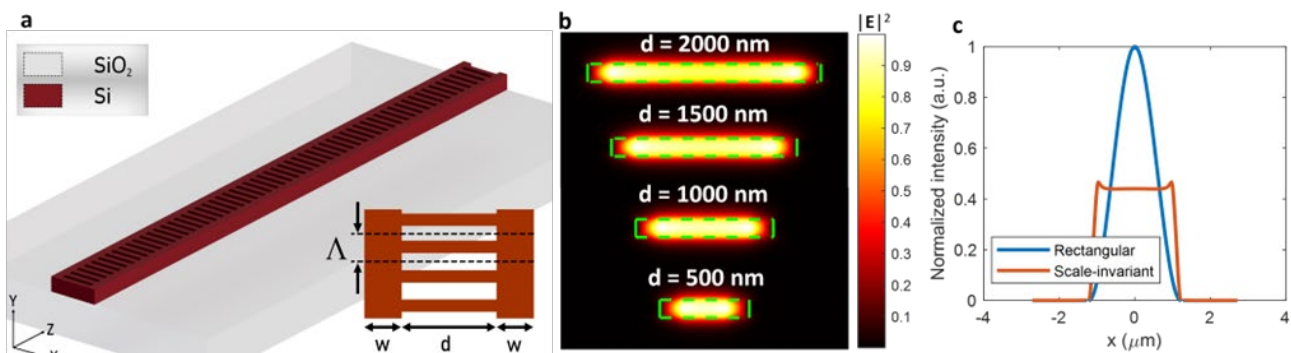

Fig. 1. (a) Schematic of the planar scale-invariant waveguide with light confinement in air, implemented on a 220 nm-thick SOI platform. The structure consists of a fully etched SWG (metamaterial) of width $d$ and period $\Lambda$ that is surrounded by two rectangular Si ridges of width $w$. The final dimensions for our design are $w$ = 217 nm, $\Lambda$ = 200 nm and a duty cycle of 0.5. (b) Simulated intensity profiles of the waveguide shown in (a) for varying values of $d$, demonstrating a uniform field profile. (c) Normalized field intensity along the $x$-axis at $y$ = 0, (defined in (a)) for the scale-invariant waveguide (red curve) and a standard rectangular waveguide of equivalent size (blue curve). The uniformity reduces maximum field intensity (at the same total power) by a factor of 2.17, enabling higher power handling.

Figure 1 shows the schematic of the scale-invariant waveguide as the building block of the structures in this paper. The design follows the concept introduced in Ref. [12], where two high-index slabs of thickness $w$ and refractive index $n_H$ are separated by an intermediate-index layer of thickness $d$ and refractive index $n_S$, and surrounded by a lower-index cladding $n_L$ ($n_H > n_S > n_L$). When the effective index of the guided mode is engineered to match $n_S$, the structure becomes scale-invariant. Thus, variations in the separation $d$ do not change the effective index of the guided mode. However, the main drawback of that configuration is its multi-material and non-planar geometry [12], which complicates fabrication and limits integration with standard photonic platforms. These limitations are avoided in the structure in Fig. 1(a), which is fabricated on a silicon-on-insulator (SOI) platform with a 220 nm-thick Si layer over a 2-μm thick $SiO_2$, with air as the top cladding. As illustrated in Fig. 1(a), our structure consists of two Si ridges of equal width $w$, separated by a SWG metamaterial region. The SWG period $\Lambda$ is chosen such that the effective index of metamaterial is equal to that of a Si waveguide of width $2w$ (the case where $d = 0$) at wavelength $\lambda = 1550$. For simplicity the duty cycle is fixed at 50%. We did not include the Si thickness as a design parameter due to commercial availability of SOI wafers. To determine the effective index of SWG, we first simulate an infinitely wide metamaterial region ($d \to \infty$). A band structure analysis is performed using the three-dimensional finite-difference time-domain (3D FDTD, Lumerical) technique with Bloch boundary conditions along the Z direction in Fig. 1(a) with the unit-cell length of $\Lambda$. Although both TE-like and TM-like modes (i.e., electric and magnetic field in the plane of the waveguide, respectively) can, in principle, be supported in this geometry, we limit our study in this paper to TE-like modes.

For a given wavevector $k_z$, the resonance frequency $\omega_{res}$ is obtained, and the effective index is calculated as

$$n_{eff} = \frac{ck_z}{\omega_{res}} \quad (1)$$

where $c$ is the speed of light. In our case, this yields $n_{eff} = 2.044$. We then calculate the width of a rectangular Si waveguide with the same effective index, obtaining 390 nm. For $d = 0$, the total width is equal to $2w$, giving $w \approx 195$ nm. Fine-tuning this width during the design gave a final $w = 217$ nm. The intensity profile of the waveguide for different values of $d$ is shown in Fig. 1(b). The normalized field intensity at $y = 0$ as a function of $x$ is plotted in Fig. 1(c) for both the scale-invariant waveguide and a conventional Si ridge waveguide with the same overall width. Figures 1(b) and 1(c) clearly demonstrate the scale-invariant property and uniform field distribution in the air region for our structure. The simulation results in Fig. 1(c) reveal that the maximum field intensity in the scale-invariant waveguide is smaller by a factor of 2.17 compared to that in a rectangular waveguide of the same dimensions, while both structures carry the same optical power. This reduction in peak intensity allows the waveguide to handle higher power without damaging the material or inducing unwanted nonlinear effects [21].

For experimental demonstration, the effective index is extracted from the transmission spectrum of an unbalanced Mach–Zehnder interferometer (MZI). In an MZI, constructive interference occurs when the phase difference between the two arms is an even multiple of $\pi$, giving the effective index as

$$n_{eff} = \frac{m\lambda}{\Delta L} \quad (2)$$

where $m$ is an integer, $\lambda$ is the wavelength, and $\Delta L$ is the path length difference of the two MZI arms [19]. For our design at λ

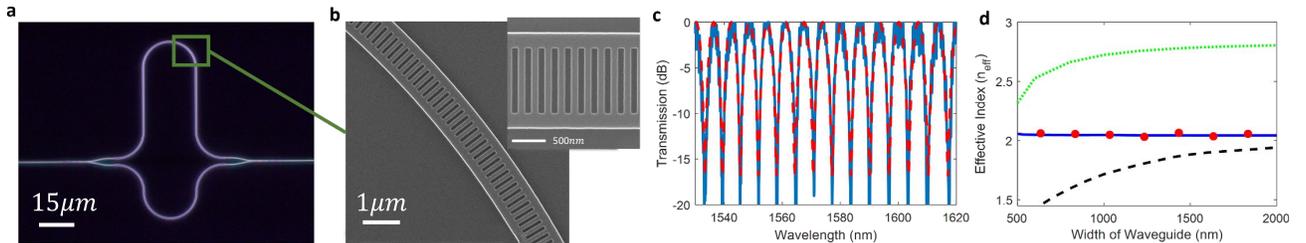

Fig. 2. (a) Optical microscope image of a fabricated MZI used to measure the effective index and the group index of the scale-invariant waveguide. (b) SEMs providing detailed views of the fabricated scale-invariant waveguide structure, highlighting the subwavelength metamaterial design. (c) Measured transmission spectrum (blue curve) of the MZI in (a) compared to the theoretical fit (red curve), showing good agreement. (d) Effective index of the fundamental guided mode as a function of the waveguide width (*2w+d*), calculated both numerically (blue solid line) using 3D FDTD simulations and experimentally (red dots) using the MZI characterization results. The results confirm the scale-invariance of the structure as the effective index remaining nearly constant across varying widths. Theoretical effective index of a rectangular Si waveguide (green dotted curve) and a SWG waveguide (black dashed curve) on the same SOI platform are also plotted for comparison.

= 1550 nm, 3D FDTD-predicted effective indices are used to determine the required $\Delta L$ values. Two unbalanced MZIs with $\Delta L$ = 12.4 µm and 12.8 µm, corresponding to $m$ = 16 and $m$ = 17, respectively, are designed and fabricated such that each produces a single constructive-interference peak within the measurement bandwidth. These peaks enable extraction of the effective refractive index of the waveguide. In addition, a third MZI with $\Delta L$ = 118.4 µm (corresponding to $m$ = 155) is implemented to estimate the group index, following the method described in Ref. [22]. All devices are fabricated on a 220-nm SOI platform using electron-beam lithography (EBL) and dry etching. Hydrogen silsesquioxane (HSQ) resist is deposited and patterned by 100 kV EBL (Elionix) with ESpacer charge dissipation, developed in tetramethylammonium hydroxide (TMAH) and transferred into Si using inductively coupled plasma reactive-ion etching (ICP-RIE). Residual resist is removed after etching. The dark field optical microscope image of a fabricated MZI is shown in Fig. 2(a). Fig. 2(b) shows the scanning electron microscopy (SEM) images of a straight and a curved scale-invariant waveguide used in the MZI structure. Transmission spectra of the fabricated MZIs are measured using a tunable laser and a high-speed spectrum analyzer, and fitted by least-squares regression in MATLAB, as shown in Fig. 2(c). The measured transmission spectrum (blue) closely matches the theoretical fit (red) obtained from the MZI model. The extracted effective indices for different $d$ are presented in Fig. 2(d), where the total waveguide width is defined as $2w+d$. The nearly constant values of the effective index for different widths and the good agreement between simulation (blue line) and experimental (red dots) results confirm the scale-invariance of the fabricated structures. For comparison, Fig. 2(d) also includes the effective indices of two reference structures with the same total width $(2w + d)$ on the same SOI platform; the same incremental changes in this total width are applied to all three cases. The first reference is a conventional Si ridge waveguide with width $2w + d$ (green dashed), and the second is a pure SWG waveguide (black dashed) designed with the same total width, the same incremental width steps, and the same grating period and duty cycle as the scale-invariant waveguide. As theoretically expected, the scale-invariant index exhibits a transition from conventional-waveguide behavior at small widths to sub-wavelength grating behavior at large widths. Finally, higher-order MZI measurements indicate a group index between 3.2 and 3.8, confirming the absence of slow-light behavior.

To study the operation bandwidth of the scale-invariant waveguide, we calculate its fundamental TE-like mode profile over the 200 nm wavelength range at three distinct wavelengths: 1450 nm, 1550 nm and 1650 nm (see Fig. 3). The field remains uniform for this wavelength range (less than ±6% deviation compared to the intensity at the center wavelength), demonstrating a large bandwidth for the scale-invariant operation. Our simulations show stronger deterioration of the field uniformity beyond the 200 nm bandwidth. Nevertheless, we did not perform a systematic optimization to maximize the bandwidth beyond 200 nm. Nevertheless, to the best of our knowledge, this is the largest bandwidth reported for any scale-invariant SOI waveguide at telecommunication wavelengths.

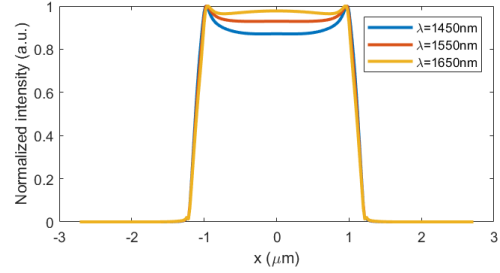

Fig. 3. Normalized electric field intensity along the width of the waveguide for 1450 nm, 1550 nm and 1650 nm wavelengths.

Another important consideration in the design of all sensitive structures is the sensitivity to fabrication imperfections. We analyze the effect of fabrication imperfections on the performance of the designed scale-invariant waveguide by applying a set of error models to the key geometric parameters of the device in Fig. 1(a). Fabrication errors are assumed to follow a normal distribution centered at zero, with a standard deviation $\sigma = \frac{\text{maximum error}}{3}$. This assumption implies that the specified maximum error corresponds roughly to three standard deviations, covering 99.7% of the distribution according to the empirical rule. The maximum error is obtained based on the limitations of the fabrication process and its corresponding tools. Figure 4(a) shows the variation of the Si device layer thickness from 200 nm to 240 nm, which represents the extreme deviation of ~ ±10% in the Si thickness (typical Si device-layer variations in commercial SOI wafers are < 5%). Figure 4(b) shows the normalized electric field distribution across the waveguide for different thicknesses in Fig. 4(a), clearly demonstrating that the waveguide mode remains nearly uniform (scale-invariant) for different thicknesses in such a wide range. Different values of $w$ are considered in Fig. 4(c) to account for lithographic deviations introduced during fabrication. The corresponding normalized electric-field profiles of the guided mode, shown in Fig. 4(d), exhibit only minimal variation, even with a deviation of ~15 nm in $w$, which represents the upper bound of practical EBL patterning errors. Taken together, the statistical studies shown in Fig. 4 confirm that our scale-invariant waveguide exhibits minimal sensitivity to fabrication imperfections even at the highest extremes.

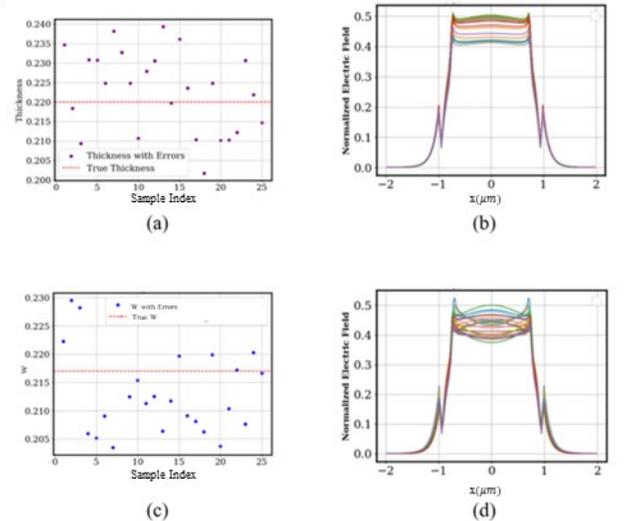

Fig. 4. Parameter perturbations used to simulate the effect of errors for the structure in Fig. 1(a): (a) Si thickness variations with respect to the designed value, (b) the normalized electric field under varying thickness (each colored curve corresponds to the Si thickness identified by the same color in (a), (c) deviations in $w$ compared to the baseline design, and (d) the corresponding normalized electric field distributions for different values of $w$.

The robust scale-invariant waveguide platform in this work can be used as the main building block to form scale-invariant resonators (e.g., by forming a ring or a racetrack) with a uniform optical mode confined within the low-index (air) region to enable interaction with a large material volume (e.g., in gas and biosensing). This contrasts with conventional ridge or slot resonators, where the field is localized in narrow regions. To realize this concept, we design and fabricate racetrack resonators based on the scale-invariant waveguide. The resonator is coupled to a scale-invariant waveguide with similar properties as shown in Fig. 5(a). The coupling condition between the waveguide and the resonator is optimized using numerical simulations. Because the optical mode is strongly confined in the $x$ direction (Fig. 1(b)), direct coupling to the adjacent waveguide is inherently weak. Therefore, racetrack geometries are required to provide sufficiently long interaction lengths. The coupling behavior of the scale-invariant racetrack resonator is analyzed using coupled-mode theory, where the power coupled between two identical waveguides $P_c$ is given by

$$P_c = sin^2[\kappa(L + L_0)] \quad (3)$$

where $\kappa$ is the coupling coefficient, and $L$ is the physical coupling length. Because the resonator–waveguide coupling region exhibits nonzero modal overlap even at $L = 0$, an effective offset length $L_0$ was introduced and extracted from short-length electromagnetic simulations. The loaded Q was evaluated using:

$$Q = \frac{2\pi n_g L_t |\tau| e^{-\alpha L_t/2}}{\lambda(1 - |\tau|^2 e^{-\alpha L_t})}, \quad |\tau| = \sqrt{1 - P_c} \quad (4)$$

where $L_t$ is the total racetrack length, $\alpha$ is the attenuation coefficient, and $n_g = 3.4$ is the group index. 3D FDTD simulations are used to estimate the coupling strength for practical device geometries. For a nominal coupling gap of 200 nm and waveguide parameter $d = 1\ \mu m$, the simulations yield $\kappa$=0.01 $\mu m^{-1}$, with an effective offset length of $L_0 = 5.1\ \mu m$. Based on these parameters, a coupling length of approximately 70 $\mu m$ is selected for fabrication, for a racetrack of total length $L_t = 321\ \mu m$. This choice ensures that sweeping the coupling gap across fabrication covers a wide range of coupling regimes. The fabrication process is optimized by implementing a multipass EBL writing strategy to minimize patterning noise and sidewall roughness, ultimately enabling the higher values of Q [23]. Optical and SEM images of the scale-invariant racetrack resonators with the smallest fabricated gap of 110 nm (over-coupled operation) are shown in Figs. 5(a) and 5(b), respectively. The measured transmission spectrum (Fig. 5(c)) at critical coupling with a gap of 290 nm exhibits a loaded Q of 20000, corresponding to an intrinsic Q ~ 40000 and a waveguide loss of $\alpha \approx$ 1.5 dB/mm. This is close to the best values reported for SWG resonators on the same platform [24] with a nonuniform field distribution. The resulting critically coupled scale-invariant resonator with intrinsic Q ~ 40,000 and strong light confinement in air can play a major role in future light-matter interaction applications. Future efforts will be focused on further improvement of the fabrication quality as well as new designs for enabling single-mode operation at higher waveguide/resonator widths.

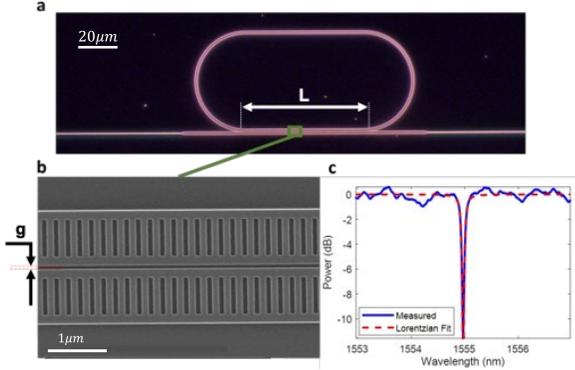

Fig.5. (a) Optical microscope image of the fabricated scale-invariant racetrack resonator coupled to a scale-invariant waveguide. (b) SEM image of the smallest fabricated gap in the over-coupled configuration. (c) Measured transmission response of the racetrack resonator (blue curve) with g = 200 nm, d = 1 μm and its Lorentzian fit (red curve), showing a loaded Q of 20000.

In summary, we demonstrated here a planar metamaterial-based scale-invariant waveguide on an SOI platform for realization of resonators with strong field confinement in air, large interaction volumes with unform light intensity, and reasonably high intrinsic Q ~ 40,000. By embedding a sub-wavelength periodic metamaterial region within a single-material Si waveguide, we achieved scale-invariance in the planar direction while maintaining compatibility with standard CMOS fabrication processes. Through numerical simulations and experimental measurements, we verified the scale-invariance of the designed structures with robustness against fabrication imperfections. The strong field localization within the air region makes the structure highly suitable for light-matter interaction applications like sensing, quantum photonics, and nonlinear optics. Beyond light-matter interaction, the ability to engineer scale-invariant waveguides with uniform field profiles (and no hot spots) offers significant advantages, including enhanced power handling and robustness to fabrication imperfections. Future efforts will focus on further optimizing the design for enhanced performance and expanding its applicability to other material platforms and operation wavelengths.